# A new perspective on Digital Twins: Imparting intelligence and agency to entities

Ashwin Agrawal
Department of Civil and Env. Engineering
Stanford University
Stanford, USA
ashwin15@stanford.edu

Vishal Singh
Centre of Product Design and Manufacturing
Indian Institute of Science
Bangalore, India
singhv@iisc.ac.in

Martin Fischer
Department of Civil and Env. Engineering
Stanford University
Stanford, USA
fischer@stanford.edu

*Abstract*— **Despite the Digital Twin (DT) concept being in the industry for a long time, it remains ambiguous, unable to differentiate itself from information models, general computing, and simulation technologies. Part of this confusion stems from previous studies overlooking the DT's bidirectional nature, that enables the shift of agency (delegating control) from humans to physical elements, something that was not possible with earlier technologies. Thus, we present DTs in a new light by viewing them as a means of imparting intelligence and agency to entities, emphasizing that DTs are not just expert-centric tools but are active systems that extend the capabilities of the entities being twinned. This new perspective on DTs can help reduce confusion and humanize the concept by starting discussions about how intelligent a DT should be, and its roles and responsibilities, as well as setting a long-term direction for DTs.**

## I. Introduction

Digital Twin (DT) has emerged as a key concept in the digital transformation of the industry. It was originally described as a digital information construct of a physical system, linked with the physical system in question. DT consists of three parts [1]: (1) the physical entity, (2) the virtual model, and (3) the bi-directional data flow between them. Any change in the state of the physical object leads to a change in the state of the virtual model and vice-versa.

However, despite the DT concept being in the industry for about twenty years, it is still ambiguous [2], not being able to distinguish itself from general computing and simulation technologies [3]. For some people, the term DT might imply a highly sophisticated model with predictive and prescriptive capabilities [4], [5]. And for others, DT might just be a simple digital representation [6], [7]. [8], [9] reviews the major DT definitions existing in practice and finds different perspectives. Some state DT is a software representation, or a digital model, or a Cyber-physical system. There is also confusion about whether DT is a replica of a product, process, or a system.

The prominence of the problem in practice is further confirmed by using a web-based tool, Answer the Public [10], that analyzes the Google search trends for the keyword "Digital Twin." Fig. 1 groups the results into three categories. The fuzziness around the concept of DT is quite evident from the most common queries listed under each category including "What is a true DT," "DT v/s Industry 4.0," "DT v/s cyber physical systems," and "DT and AI."

The usage of DT across such a wide range of things and applications risks it being rejected by the people due to its vagueness and as hype [2]. It also results in practitioners getting confused while deploying DTs [11], setting unrealistic expectations [12], and misaligning on strategic decisions [13]. To alleviate this challenge, [9], [14] calls for a fresh perspective to unify different existing perspectives of DTs. Similarly, [8] argues that a new viewpoint on DTs is needed that emphasizes on the purpose of DT rather than a general representation.

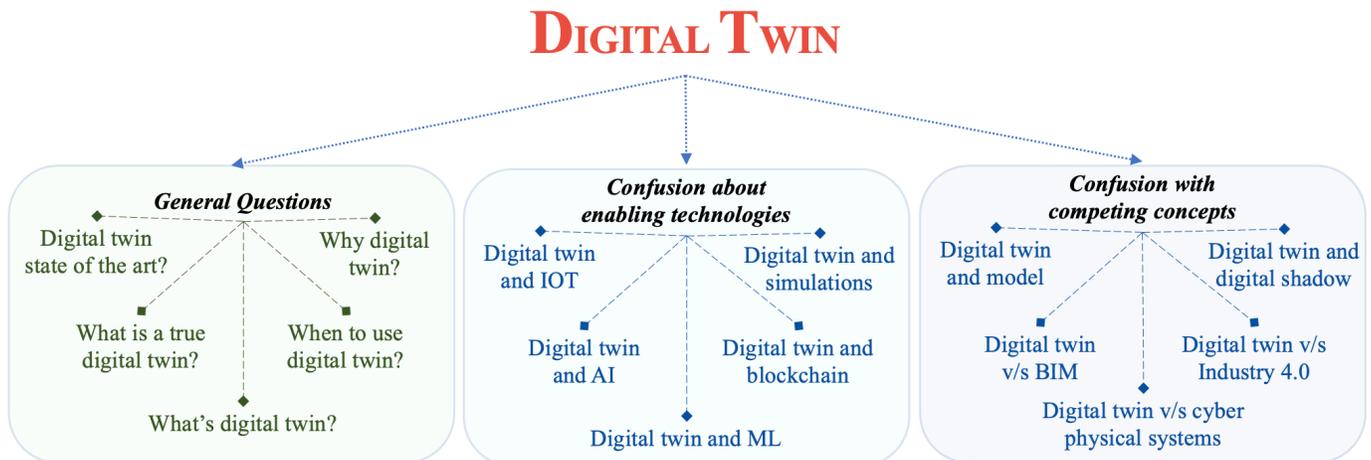

Figure 1 Scrapped data about Digital Twin from Google Search clustered into three categories

This work is funded by Center of Integrated Facility Engineering (CIFE), Stanford University

This paper thus presents a new perspective on DTs that has emerged from: (1) Detailed observations and 35+ hours of expert interviews from the ongoing longitudinal study that has examined DT implementations since 2018 [11], [14]; and (2) reflections based on observing more than 1000 DT implementations since 2008 as a part of Stanford Virtual Design and Construction (VDC) certificate program [15], [16]. The digital trail of the longitudinal study can be found at [17], [18]. The contribution of this paper is twofold: (1) A new perspective for understanding DTs based on empirical observations is presented, and (2) A case is made on how the new perspective can aid future research and alleviate some of the existing confusion around the DT concept.

This work is organized as follows: in section II we introduce the new proposed definition for DT. The benefits of this new perspective are discussed in section III. Section IV concludes the paper and Section V discusses the limitations and future research.

## II. New Perspective For Understanding Digital Twins

**Proposed definition:** "A digital twin is a digital replica that imparts some form of <u>intelligence</u> and <u>agency</u> into the <u>entities</u> being twinned to achieve the <u>desired function</u>"

### A. Intentional use of the word 'entities'

The researchers observed that even though most practitioners wanted to digitally twin a physical product, sometimes they questioned why their process or organization cannot be digitally twinned. For example, one of the practitioners questioned can they digitally twin their project organization to determine where the bottlenecks in communication are? [19] also showcases a DT of the organization that allows scenarios to be thoroughly tested to predict the performance of potential tactics and strategies.

We found that most of the literature on DT has focused on DT as a digital replica of tangible physical assets [17]. However, people have recently started realizing that focusing only on the physical product can limit the application of a DT. Therefore, broader definitions have emerged that include twinning the processes, the systems, and the organization. For example, General Electric (GE) defines DT as a software representation of a physical asset, system or a process [20], and PwC puts DT as a way to capture the virtual model of the organization [21]. Gartner's states that DT can be a digital representation of a physical object, process, organization, person, or other abstractions [22]. To indicate the stance that a DT can be a replica of any entity depending on the task in hand, the authors deliberately refrain from defining DT as a replica of physical asset, product, process, and organization.

### B. Emphasis on the 'desired function'

The researchers were often asked by the practitioners what capabilities a DT should have. One of the participants asked if they need to have real-time data flow even if they do not need it for their use case? Another participant asked why do they need to have Artificial Intelligence (AI) (suggested a necessary component of DT by [23]) in a DT if they are not using predictive analytics at all?

Therefore, we argue that there is no so-called universal DT that everyone can deploy. DT's can have a wide variety of sophisticated capabilities, ranging from simple digital representation to increasingly complex models with predictive and prescriptive capabilities [11]. Lack of research in this field has resulted in the rebranding and reuse of emerging technological capabilities such as prediction, simulation, AI, and Machine Learning (ML) as necessary constituents of DT [24]. By forcing specific technologies into every use case and not having a fit-for-purpose DT can result in biases towards a particular technology and therefore result in missed opportunities [14]. Hence, we have deliberately avoided listing any technical capabilities required in a DT, since we believe the capabilities should follow the desired function for a specific use case, as also noted by [8].

### C. Introduction of 'intelligence' and 'agency'

The digital format of the data, which for long existed in physical (analog) form, can now be sensed, interpreted, analyzed, and acted upon by a computer (virtual model), like the way humans' function. For instance, like human sense organs, DT uses sensors, which helps it perceive the environment by enabling data flow from the entity to the virtual model. The virtual model, like the human brain, enables information processing and decision making. Finally, the data flow from the virtual model to the entity through actuators (bidirectionality of DTs) enables a DT to perform actions, like human muscles. Therefore, in a way, the entity being twinned gets an "artificial brain", providing it the possibility to be intelligent and act autonomously.

To define 'intelligence', we adapt Alan Turing's operational definition [25], which is to achieve human-level performance in cognitive tasks. For DTs to achieve human-level intelligence, they need to master many cognitive aspects, which, of course, is not a one-time development task. This is where the long-term vision of what an ideal (intelligent) DT might look like becomes vital to guide the development of DTs in the industry. Our perspective forces practitioners and researchers to start a discussion about this vision as discussed in Section 3. 'Agency' is defined as level of control over final action by DTs.

Fig. 2 describes an example to further explain the above concept. In the example, a building's temperature rises beyond ambient levels, causing user discomfort. Humans inform the operator about the problem in the baseline scenario. The operator decides how much cooling to increase based on the time of day, number of occupants, and energy consumption, and the operator then adjusts the temperature accordingly. If there is a one-way data flow from the entity to the virtual model, as in the second scenario, sensors can detect an increase in temperature and the virtual model can calculate how much cooling load needs to be increased, reducing the need for a human operator. Therefore, in a way, the building becomes somewhat intelligent as it can take decisions on its own. However, an operator still needs to implement the decision, since there is no data flow from the virtual model to the entity, and therefore no agency. The agency finally gets added in the third scenario, where the bi-directional data flow (DT's basic requirement) enables the computer (virtual model or the "artificial brain") to implement the decision using actuators.

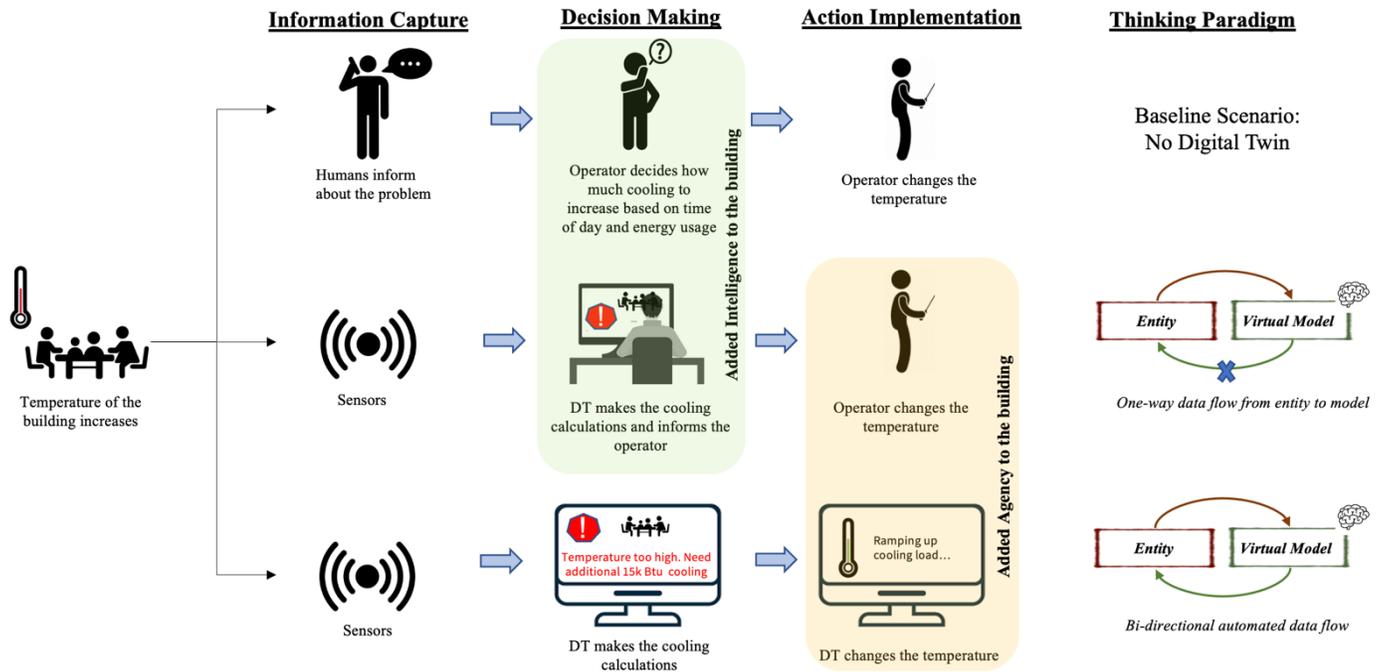

*Figure 2 A real world example showcasing how DT helps the entity being twinned digitally to gain intelligence and agency*

Our goal is not to advocate for complete agency or intelligence of DTs. We highlight the example in Fig. 2 to give an intuition to the readers of the concept of agency (and intelligence) and also illustrate the extremes where DTs work directly without being supervised by humans. Of course, in the real world, there would be many intermediate cases of agency (and intelligence) where the DT would only independently handle certain scenarios, or a human would be involved in the loop. For example, in some cases like disaster responses, building humans in the digital twin is essential, and therefore the data would flow from virtual model to human and then from human to entity.

### III. BENEFITS OF THE NEW PERSPECTIVE

#### A. Alleviates the existing confusion around DT

The new viewpoint promotes the idea that a DT does not have to be all-or-nothing; it may range along a spectrum of intelligence and autonomy (agency) capabilities, with some DTs being more intelligent and having more agency than others as per the required function. The confusion around the various definitions of DT proposed by different researchers alleviates once we look at them through this new perspective. Readers should realize that none of the definitions of DT offered by researchers is incorrect, but rather each definition includes only a partial list of intelligence abilities in a DT, that makes sense for specific use cases. For example, some definitions put an emphasis on basic intelligence abilities like data manipulation (therefore emphasizing digital representation) while others place emphasis on advanced intelligence abilities (emphasizing prediction and prescription). Future research can explore different levels of intelligence capabilities in a DT (similar to levels of automation [26]) to facilitate clear communication between stakeholders and enable practitioners to select a fit-for-purpose DT and have a more informed conversation about the question: "How intelligent DT do I need?".

In addition, this new perspective also differentiates DT from general computing and simulation technology by emphasizing the concept of agency that is brought about by bi-directional data flow in DTs. Unlike general simulation technologies, DTs are not just expert-centric tools. In fact, DTs can work by themselves (agency) to make decisions, and adapt themselves and the entities being twinned to achieve the desired goals without human intervention [27], as shown in Fig.2 (case-3).

#### B. Makes the concept of DT more relatable and intuitive

The new perspective humanizes the DT concept, making it more relatable and intuitive by making it technology agnostic, emphasizing that all the technologies are simply a means to achieve the end goal of intelligence and agency. For example, it becomes very natural to ask and comprehend questions such as: "What is the ideal level of intelligence in a DT," "What roles can a DT perform," "Who is responsible if a DT makes mistakes," and "What kind of situations can a DT handle."

We believe that viewing the concept of DT as a mere digital replica undermines the possibilities that the technology can offer. The new perspective presented about the DT concept, focuses on the end goal, which we believe is to impart some form of intelligence and agency into the rather "dumb" entities by means of an artificial brain (in our case the DT). On a similar line, [27] also notes that with the increased capabilities and autonomy, DTs would not merely be an expert-centric tool but proactively make decisions, complete tasks, and adapt to changing conditions. Future research can explore how DTs can partner with humans in a work system, developing models for work delegation between the two, and investigating trust privacy, and ethics.

*C. Helps setting up a long term vision for DT*

Creating the most advanced DT for practice is an evolving process and not a one-shot task [24]. To ensure smooth deployment of DT in the industry, practitioners need to start small and make steady progress towards the envisioned 'ideal' DT. However, the term 'ideal' DT is abstract and currently there are no clear guidelines on what an 'ideal' DT looks like [28]. This risks practitioner going with their "best-suited option" and thus selecting a sub-optimal or an overly optimistic solution that might later have to be altered or disposed-off completely [28].

The new perspective presented can help alleviate the above problem to some extent. For example, the idea of intelligence and agency opens up avenues for future research to create a 'Turing test' [25] for DTs. This can help research community and practitioners understand what an ideal DT might look like and work towards it, similar to how Alan Turing described what did it mean for a computer to be intelligent.

## IV. CONCLUSION

Adoption of DTs in practice would be largely determined by how it gets interpreted by practitioners. We find that the multiple existing definitions and perspectives around the DT concept is confusing practitioners, thus risking its rejection as hype. Therefore, to alleviate the challenge, this work proposes a new perspective to understand DTs based on empirical observations from a longitudinal study. This new perspective can help unify the existing perspectives, humanize the concept by starting discussions about how intelligent a DT should be, and its roles and responsibilities, as well as setting a long-term direction for DTs.

## V. LIMITATION AND FUTURE RESEARCH

The proposed definition (and perspective) of DT is primarily based on the DT case-studies conducted in the building and construction sector. Therefore, the generality and applicability of this definition cannot be claimed outside the building and construction sector. However, based on the experience and conversations with the industry professionals from manufacturing, aerospace, and healthcare sector, the authors anticipate similar findings in other sectors. Future research can explore the universality of this perspective.

ACKNOWLEDGMENT

We would like to thank all the industry and academic experts who gave us their valuable time and feedback. We acknowledge the financial support provided by the Center for Integrated Facility Engineering (CIFE) at Stanford University for completing this project. The authors would also take this opportunity to express their gratitude to Rui Liu, Stephanie Chin, Melody Spradlin, Cynthia Brosque, Simge Girgin, and Alberto Tono for their valuable feedback. The authors also thank the conference reviewers who gave us their valuable feedback on the paper. Last but not least, the first author would also like to especially thank Tulika Majumdar, Alissa Cooperman, and Hesam Hamledari for their invaluable feedback and help during the whole research process.